\documentclass[conference,peer-reviewed]{aesconf}
\usepackage{amsmath,graphicx}

\usepackage{subfigure} 
\graphicspath{{./image/}}

\usepackage{amssymb}

\usepackage{epstopdf}
\usepackage{tabularx}
\usepackage{multirow}
\usepackage{textcomp}
\usepackage{url}
\usepackage{lipsum}
\usepackage{makecell}

\usepackage{bm}
\usepackage{calc}

\usepackage{color}
\usepackage{dsfont}

\usepackage[utf8]{inputenc}

\usepackage{microtype}

\usepackage[numbers,square]{natbib}

\usepackage{booktabs}

\usepackage{url}

\title{Beyond Equal-Length Snippets: How Long is Sufficient to Recognize an Audio Scene?}

\author[1,2]{Huy Phan}
\author[2]{Oliver Y. Ch\'{e}n}
\author[3]{Philipp Koch}
\author[1]{Lam Pham}
\author[1]{Ian McLoughlin}
\author[3]{Alfred Mertins}
\author[2]{Maarten De Vos}

\affil[1]{University of Kent, School of Computing, UK}
\affil[2]{University of Oxford, Department of Engineering Science, UK}
\affil[3]{University of L\"ubeck, Institute for Signal Processing, Germany}

\correspondence{Huy Phan}{h.phan@kent.ac.uk}

\lastnames{Phan et al.}

\shorttitle{How Long is Sufficient to Recognize an Audio Scene?}

\begin{document}

\twocolumn[
\maketitle 

\begin{onecolabstract}
Due to the variability in characteristics of audio scenes, some scenes can naturally be recognized earlier than others. In this work, rather than using equal-length snippets for all scene categories, as is common in the literature, we study to which temporal extent an audio scene can be reliably recognized given state-of-the-art models. Moreover, as model fusion with deep network ensemble is prevalent in audio scene classification, we further study whether, and if so, when model fusion is necessary for this task. To achieve these goals, we employ two single-network systems relying on a convolutional neural network and a recurrent neural network for classification as well as early fusion and late fusion of these networks. Experimental results on the LITIS-Rouen dataset show that some scenes can be reliably recognized with a few seconds while other scenes require significantly longer durations. In addition, model fusion is shown to be the most beneficial when the signal length is short.
\end{onecolabstract}
]

\section{Introduction}
\label{sec:intro}

Audio scene recognition (ASC) \cite{Stowell2015, phan2017b} is an important task in machine hearing \cite{Lyon2010, Wang2006}. A common goal in ASC is to 
use machines to recognize environments based on acoustic signals, as in human hearing. Over the past few years, research on this problem has advanced rapidly, both in performance \cite{phan2017c, Zadeh2016, Mun2017, Sakashita2018, Zhang2018} and in the number of available datasets \cite{Mesaros2018b, Mesaros2018, Mesaros2017, Rakotomamonjy2015}.

Audio scenes vary significantly in their characteristics, i.e. their background noise and foreground sound events, and, as a result, the duration required for human ears to perceive and recognize them is different from one to another. For instance, one might recognize with relative ease an ``in airplane'' acoustic environment within a few seconds due to its loud and distinguishable background noise. However, it may take minutes of listening to  accumulate sufficient acoustic cues to differentiate a ``restaurant'' from a ``cafe'' or even a ``busy street'' with similar babble noise. As one of the goals of machine hearing is to achieve human hearing intelligence, this variability should also be generalizable to a machine hearing system. However, most, if not all, available ASC datasets assume a fixed signal length for every scene category and instance. 
Typical lengths of 30 seconds \cite{Mesaros2018,Rakotomamonjy2015} or 10 seconds \cite{Mesaros2018b,Mesaros2017} have been commonly adopted when designing such datasets. There also exist studies with shorter signal duration, such as six seconds \cite{Guo2017}. Although a fixed common length makes dataset design and experimentation easier, it does not reflect the reality of the ASC task. Hence, this paper aims to investigate into the perspective, namely how long is sufficient to reliably recognize different acoustic scenes using state-of-the-art models. Findings of this study may also suggest how an ASC should be implemented in real applications to improve the quality of services. Our results show that, for such a system, the listening duration should be adapted to different kinds of scenes in order to reach a certain level of recognition certainty.

Methodology-wise, approaches based on convolutional neural networks (CNNs) \cite{Zadeh2016, Valenti2016, phan2017b, phan2017a, Mun2017} and recurrent neural networks (RNNs) \cite{phan2017c, Vu2016, Zhang2018} have been demonstrated to be most efficient for ASC. In addition, state-of-the-art performance has often been achieved with ensembles of multiple networks \cite{Sakashita2018, Bae2016, Yin2018}. Although ensemble methods have been well-established to improve performance of machine learning systems, and have been applied to ASC as a rule of thumb, no prior work has studied whether and to which temporal extent model fusion is necessary and most useful for the ASC task. We therefore aim to examine this question in this work. To accomplish these goals, we use two state-of-the-art single-network models based on CNN and RNN as well as two fusion schemes of these models for classification purpose. An early fusion scheme constructs a two-stream convolutional-recurrent neural network (C-RNN)  and explores in-network fusion of the two streams' learned features before classification. A late fusion scheme trains two standalone networks independently and probabilistically aggregates their classification results afterwards. The experiments are based on the LITIS-Rouen dataset \cite{Rakotomamonjy2015}. 


\section{CNN and RNN for ASC}

The CNN model proposed here improves on its counterpart in \cite{phan2017b} in such a way that the convolutional filters are trained to take into account invariance across multiple input feature channels. 
The RNN model is based on~\cite{phan2017c}, which reported  state-of-the-art performance on the LITIS-Rouen dataset \cite{Rakotomamonjy2015}. Both the CNN and RNN models share common input features.

\subsection{Input features}

An audio signal, e.g. a segment of 4 seconds long (see more detail in Section \ref{ssec:dataset}), is first decomposed into small frames of 250\,ms with 50\% overlap. We employed the label tree embedding (LTE) features proposed in \cite{phan2016d} to represent an audio frame.
To this end, low-level feature vectors are first extracted for the audio frames and used to construct a label tree. 
Given $C$ scene categories, the constructed label tree consists of $C-1$ split nodes which index $2(C-1)$ meta-classes in their left and right child nodes. 
A low-level feature vector of an audio frame is then mapped into an LTE feature vector $\mathbf{x} \in [0,1]^F$, $F = 2(C-1)$, whose entries encode the posterior probabilities of the audio frame belonging to the meta-classes. 
As a result, we obtain $T$ such LTE feature vectors for the audio snippet consisting of $T$ frames.

We employ three different low-level feature sets: (1) Gammatone spectral coefficients \cite{Ellis2009}, (2) MFCCs, and (3) log-frequency filter bank coefficients. 
We extract LTE features with the presence/absence of the background noise as in \cite{phan2017c,phan2017b}. 
In total, we obtain $D=6$ feature channels so the audio snippet is eventually represented by a multi-channel image feature $\mathbf{X} \in [0,1]^{F \times T \times D}$.

\subsection{CNN}
\label{ssec:cnn}
Audio scene recognition using CNNs on LTE features have been previously explored in \cite{phan2017b}. In that work, over-time convolution with 3-dimensional convolutional filters, which fully cover feature and channel dimensions \cite{phan2017b}, did not explore the feature invariance across LTE channels. As an improvement, the CNN proposed here is designed to have 2-dimensional convolutional filters to perform convolution over time as well as across input feature channels.

Let us denote such a 2-dimensional filter as $\mathbf{w} \in \mathbb{R}^{F \times w}$ where $w < T$ represents the temporal width of the filter. Convolving the filter $\mathbf{w}$ with the multi-channel image input $\mathbf{X}$ over-time and across-channel results in a 2-dimensional feature map $\mathbf{O} \in \mathbb{R}^{(T-w+1) \times D}$ whose entries are given by
\begin{align}
	o_{ij} = (\mathbf{X} * \mathbf{w})_{ij} = \sum_{m,n}(\mathbf{X}_{j}[i : i + w - 1] \odot \mathbf{w})_{m,n}.
\end{align}
Here, $*$ and $\odot$ indicate the convolution and element-wise multiplication operations, respectively. $\mathbf{X}_{j}[i : i + w - 1]$ denotes an image slice from time index $i$ to $i + w - 1$ on the channel index $j$.  \emph{Rectified Linear Units} (ReLU) activation \cite{Glorot2011} is then applied, followed by 1-max pooling \cite{Kim2014,phan2016c} on the 2D feature map to retain the most prominent feature:
\begin{align}
	z^{\text{conv}} = \max_{i,j }o_{ij}.
\end{align}

Similar to \cite{phan2017b}, we design the CNN to have $R=3$ filter sets corresponding to three temporal widths $w \in \{3,5,7\}$ with each filter set consisting of $Q$ convolutional filters with the same width. We adopt $Q=1000$ as a large number of convolutional filters was shown to be more efficient for this ``shollow'' and simple CNN \cite{phan2017b}.
The total number of filters is, therefore, $R \times Q$; this leads to a convolutional feature vector $\mathbf{z}^{\text{conv}} \in \mathbb{R}^{RQ}$ after the pooling layer.

During network training, the convolutional feature vector $\mathbf{z}^{\text{conv}}$ is presented to a softmax layer for classification. The CNN is trained to minimize the cross-entropy error over the training examples:
\begin{align}
	E(\boldsymbol{\theta}) = -\sum_{n}{\mathbf{y}_n\log\big(\mathbf{\hat{y}}_n(\mathbf{X}_n\,,\boldsymbol{\theta})\big)} + \frac{\lambda}{2}\left\|\boldsymbol{\theta}\right\|^2_2.
	\label{eq:cross_entropy}
\end{align}
Here, $\boldsymbol{\theta}$ denotes the network parameters and the $\lambda$ denotes the hyper-parameter of the {$\ell_2$-norm} regularization term. For further regularization, dropout \cite{Srivastava2014} is also applied to the CNN feature map.

After network training, we extract the convolutional feature vector $\mathbf{z}^{\text{conv}}$ and train a linear SVM for classification, in replacement of the softmax as in \cite{phan2017b, phan2017c}.

\subsection{RNN}
\label{ssec:rnn}

For the RNN, we stack different channels of the multi-channel LTE image input $\mathbf{X}$ in the feature dimension and treat it as a temporal sequence of feature vectors $(\mathbf{\tilde{x}}_1,\mathbf{\tilde{x}}_2, \ldots, \mathbf{\tilde{x}}_T)$ where each $\mathbf{\tilde{x}}_i \in \mathbb{R}^{FD}$, $1 \le t \le T$. The RNN then reads the input sequence into the sequence of recurrent output vectors $(\mathbf{z}_1,\mathbf{z}_2, \ldots, \mathbf{z}_T)$, where
\begin{align}
	\mathbf{z}_t &= \mathbf{h}_t \mathbf{W}_{z} + \mathbf{b}_{z},
	\label{eq:rnn_output} \\
	\mathbf{h}_t &= \mathcal{H}(\mathbf{\tilde{x}}_t\,, \mathbf{h}_{t-1}).\label{eq:rnn_hidden}
\end{align}
$\mathbf{h}_t \in \mathbb{R}^H$ denotes the hidden state vector of size $H$ at time step $t$, $\mathbf{W}_{z} \in \mathbb{R}^{H \times H}$ is a weight matrix and $\mathbf{b}_{z} \in  \mathbb{R}^{H}$ denotes a bias term. $\mathcal{H}$ represents the hidden layer function of the recurrent layer and is realized using a Gated Recurrent Unit (GRU) \cite{Cho2014}. The RNN is designed to have two recurrent layers with the hidden state vector size of $H=256$. The recurrent layers are stacked on top of each other to construct a deep RNN similar to \cite{phan2017c,Graves2013}.

We eventually retain the output vector at the last time index $T$ as the recurrent feature vector, i.e. $\mathbf{z}^{\text{rec}}~\equiv~\mathbf{z}_T~\in~\mathbb{R}^H$, as it is expected to have encoded information of the entire input sequence. Dropout is also applied to the recurrent feature vector for regularization. Similarly to the CNN case, softmax and the cross-entropy loss given in (\ref{eq:cross_entropy}) are employed for classification during network training. After training, the recurrent features $\mathbf{z}^{\text{rec}}$ are also extracted and used to train a linear SVM classifier for classification.

\begin{figure} [!b]
	\centering
	\includegraphics[width=0.85\linewidth]{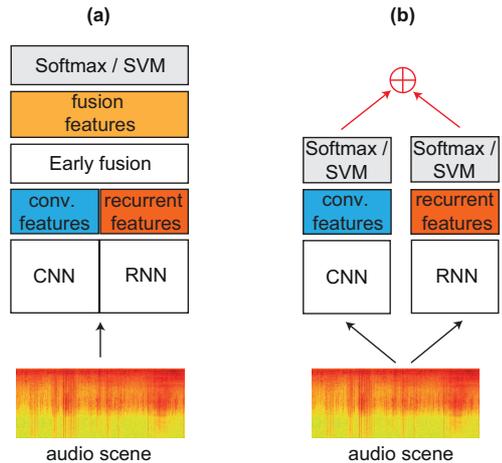}
	\caption{(a) Early fusion with two-stream C-RNN and (b) late fusion with two networks, CNN and RNN.}
	\label{fig:overview}
	\vspace{-0.2cm}
\end{figure}

\section{Early Fusion with Two-Stream C-RNN}

For the early fusion scheme, we construct a two-stream C-RNN network as illustrated in Fig. \ref{fig:overview}(a) and fuse the features learned by the CNN and RNN streams before the classification taken place.  The idea is to allow the network to explore combinations of two feature types to optimize the classification task. The CNN and RNN streams in the C-RNN network have the same body architectures as the CNN and RNN described in Sections \ref{ssec:cnn} and \ref{ssec:rnn}.

At the fusion layer of the C-RNN network, a fusion function $f~:~(\mathbf{z}^{\text{conv}}, \mathbf{z}^{\text{rec}}) \mapsto \mathbf{z}^{f}$ fuses the convolutional feature vector $\mathbf{z}^{\text{conv}}$ of the CNN stream and the recurrent feature vector $\mathbf{z}^{\text{rec}}$ of the RNN stream to produce an output map $\mathbf{z}^{f}$.  We investigate the following fusion functions $f$: \emph{sum fusion}, \emph{max fusion}, and \emph{concatenation fusion}. In addition, as $\mathbf{z}^{\text{conv}}$ and $\mathbf{z}^{\text{rec}}$ differ in size, we transform them via a fully-connected layer with \emph{sigmoid} activation beforehand to make their sizes compatible (i.e. with sum fusion and max fusion) or to equalize their contributions to the output map (i.e. with concatenation fusion);
\begin{align}
	\mathbf{\tilde{z}}^{\text{conv}} &= \mathbf{z}^{\text{conv}} \mathbf{W}^{\text{conv}} + \mathbf{b}^{\text{conv}}, \\
	\mathbf{\tilde{z}}^{\text{rec}} &= \mathbf{z}^{\text{rec}} \mathbf{W}^{\text{rec}} + \mathbf{b}^{\text{rec}}.
\end{align}
Here, $\mathbf{W}^{\text{conv}} \in \mathbb{R}^{RQ \times M}$ and $\mathbf{W}^{\text{rec}} \in \mathbb{R}^{H \times M}$ denote weight matrices while $\mathbf{b}^{\text{conv}} \in  \mathbb{R}^{M}$ and $\mathbf{b}^{\text{rec}} \in  \mathbb{R}^{M}$ denotes bias terms of the fully-connected layers. $M$ is the desired size of the transformed feature vectors $\mathbf{\tilde{z}}^{\text{conv}}$ and $\mathbf{\tilde{z}}^{\text{rec}}$.

{\bf Sum fusion.} $\mathbf{z}^{f_{\text{sum}}} = f_{\text{sum}}(\mathbf{\tilde{z}}^{\text{conv}},\mathbf{\tilde{z}}^{\text{rec}})$ computes the sum of the two feature vectors $\mathbf{z}^{f_{\text{sum}}} = \mathbf{\tilde{z}}^{\text{conv}} + \mathbf{\tilde{z}}^{\text{rec}}$.

{\bf Max fusion.} $\mathbf{z}^{f_{\text{max}}} = f_{\text{max}}(\mathbf{\tilde{z}}^{\text{conv}}, \mathbf{\tilde{z}}^{\text{rec}})$ takes the maximum of the two feature vectors $z_i^{f_{\text{max}}}\!=\! \max(\tilde{z}_i^{\text{conv}}, \tilde{z}_i^{\text{rec}})$, $1\!\le\!i\!\le\!M$.

{\bf Concatenation fusion.} $\mathbf{z}^{f_{\text{cat}}} = f_{\text{cat}}(\mathbf{\tilde{z}}^{\text{conv}}, \mathbf{\tilde{z}}^{\text{rec}})$ simply concatenates the two feature vectors to make a larger one $\mathbf{z}^{f_{\text{cat}}} = [\mathbf{\tilde{z}}^{\text{conv}}, \mathbf{\tilde{z}}^{\text{rec}}]$.

Network training and evaluation are performed similarly to the standalone CNN and RNN with a softmax layer used for classification and minimization of the cross-entropy loss given in (\ref{eq:cross_entropy}). 
In particular, dropout is further applied to the fusion features for regularization purpose. 
Again, the fusion features are extracted to train a linear SVM for classification after network training.

\section{Late Fusion of CNN and RNN}

For the late fusion scheme, we probabilistically combine the classification results of the standalone CNN and RNN in Sections \ref{ssec:cnn} and \ref{ssec:rnn}, respectively, as illustrated in Fig. \ref{fig:overview}(b). We study three fusion methods: \emph{max fusion}, \emph{mean fusion}, and \emph{multiplication fusion}. Let $\mathbf{P}^{\text{conv}} = (P^{\text{conv}}_1, P^{\text{conv}}_2, \ldots , P^{\text{conv}}_C)$ and $\mathbf{P}^{\text{rec}} = (P^{\text{rec}}_1, P^{\text{rec}}_2, \ldots , P^{\text{rec}}_C)$ denote the posterior probabilities obtained by the CNN and RNN, respectively. The classification likelihood $\mathbf{P} = (P_1, P_2, \ldots , P_C)$ after fusion with the three methods is then given by
\begin{align}
	{P_c} &= \max({P}^\text{conv}_c, {P}^\text{rec}_c) &~~\mathrm{for}~~ 1\le c \le C,& \label{eq:maximum_voting}\\
	{P_c} &= \frac{1}{2}({P}^\text{conv}_c + {P}^\text{rec}_c) &~~\mathrm{for}~~ 1\le c \le C, \label{eq:additive_voting}\\
	{P_c} &= \frac{1}{2}({P}^\text{conv}_c \times {P}^\text{rec}_c) &~~\mathrm{for}~~ 1\le c \le C, \label{eq:multiplicative_voting}
\end{align}
respectively. The final output label is then determined by likelihood maximization on the classification likelihood $\mathbf{P}$.

\setlength\tabcolsep{2.5pt} 
\begin{table*}[!t]
	\caption{Overall performance of the classification models with different test signal lengths of $\{4,8,\dots,28,30\}$ seconds. Early fusion and late fusion are abbreviated by EF and LF, respectively.}
	\footnotesize
	\vspace{-0.3cm}
	\begin{center}
		\begin{tabular}{|>{\arraybackslash}m{0.55in}|>{\centering\arraybackslash}m{0.2in}|>{\centering\arraybackslash}m{0.2in}|>{\centering\arraybackslash}m{0.2in}|>{\centering\arraybackslash}m{0.2in}|>{\centering\arraybackslash}m{0.2in}|>{\centering\arraybackslash}m{0.2in}|>{\centering\arraybackslash}m{0.2in}|>{\centering\arraybackslash}m{0.25in}|>{\centering\arraybackslash}m{0.2in}||>{\centering\arraybackslash}m{0.2in}|>{\centering\arraybackslash}m{0.2in}|>{\centering\arraybackslash}m{0.2in}|>{\centering\arraybackslash}m{0.2in}|>{\centering\arraybackslash}m{0.2in}|>{\centering\arraybackslash}m{0.2in}|>{\centering\arraybackslash}m{0.25in}|>{\centering\arraybackslash}m{0.2in}|>{\centering\arraybackslash}m{0.25in}|>{\centering\arraybackslash}m{0in}}
			\cline{1-19}
			
			\multirow{2}{*}{Method} & \multicolumn{9}{c||}{Accuracy (\%)} & \multicolumn{9}{c|}{F1-score (\%)} & \parbox{0pt}{\rule{0pt}{1ex+\baselineskip}}  \\ [0ex]  	
			\cline{2-19}
			& 4s & 8s & 12s & 16s & 20s & 24s & 28s & 30s & Avg. & 4s & 8s & 12s & 16s & 20s & 24s & 28s & 30s & Avg. & \parbox{0pt}{\rule{0pt}{1ex+\baselineskip}} \\ [0ex]  	
			\cline{1-19}
			
			CNN & $91.7$ & $94.3$ & $95.6$ & $96.0$ & $96.3$ & $96.9$ & $97.4$ & $97.5$ & $95.7$ & $91.4$ & $93.9$ & $95.3$ & $95.8$ & $96.1$ & $96.8$ & $97.3$ & $97.4$ & $95.5$ \parbox{0pt}{\rule{0pt}{.5ex+\baselineskip}} \\ [0ex]  	
			RNN & $91.2$ & $94.5$ & $95.6$ & $96.2$ & $96.6$ & $\bf 97.2$ & $\bf 97.7$ & $\bf 97.8$ & $95.8$ & $90.8$ & $94.2$ & $95.4$ & $96.0$ & $96.5$ & $\bf 97.2$ & $\bf 97.6$ & $\bf 97.7$ & $95.7$  \parbox{0pt}{\rule{0pt}{.5ex+\baselineskip}} \\ [0ex]  	
			\cline{1-19}
			EF - Sum & $90.1$ & $93.1$ & $94.5$ & $95.1$ & $95.7$ & $96.2$ & $96.5$ & $96.9$ & $94.8$ & $89.6$ & $92.8$ & $94.2$ & $94.9$ & $95.6$ & $96.1$ & $96.4$ & $96.8$ & $94.5$ \parbox{0pt}{\rule{0pt}{.5ex+\baselineskip}} \\ [0ex]  	
			EF - Max & $90.2$ & $93.1$ & $94.3$ & $94.9$ & $95.3$ & $95.9$ & $96.4$ & $96.8$ & $94.6$ & $89.5$ & $92.7$ & $94.0$ & $94.6$ & $95.1$ & $95.7$ & $96.3$ & $96.7$ & $94.3$  \parbox{0pt}{\rule{0pt}{.5ex+\baselineskip}} \\ [0ex]  	
			EF - Concat & $90.4$ & $93.1$ & $94.5$ & $95.2$ & $95.5$ & $96.1$ & $96.6$ & $96.8$ & $94.8$ & $90.0$ & $92.7$ & $94.1$ & $94.8$ & $95.3$ & $95.9$ & $96.5$ & $96.7$ & $94.5$  \parbox{0pt}{\rule{0pt}{.5ex+\baselineskip}} \\ [0ex]  	
			
			\cline{1-19}
			LF - Max & $92.0$ & $94.6$ & $95.4$ & $95.9$ & $95.9$ & $96.6$ & $96.9$ & $97.0$ & $95.5$ & $91.6$ & $94.2$ & $95.1$ & $95.7$ & $95.8$ & $96.4$ & $96.8$ & $96.9$ & $95.3$  \parbox{0pt}{\rule{0pt}{.5ex+\baselineskip}} \\ [0ex]  	
			LF - Mean & $\bf 92.1$ & $94.5$ & $95.6$ & $96.1$ & $96.5$ & $97.0$ & $97.5$ & $97.6$ & $95.9$ & $\bf 91.7$ & $94.5$ & $95.3$ & $95.9$ & $96.3$ & $96.9$ & $97.4$ & $97.6$ & $95.7$ \parbox{0pt}{\rule{0pt}{.5ex+\baselineskip}} \\ [0ex]  	
			LF - Mult & $\bf 92.1$ & $\bf 94.9$ & $\bf 95.9$ & $\bf 96.3$ & $\bf 96.7$ & $\bf 97.2$ & $\bf 97.7$ & $\bf 97.8$ & $\bf 96.1$ & $\bf 91.7$ & $\bf 94.6$ & $\bf 95.6$ & $\bf 96.2$ & $\bf 96.6$ & $97.1$ & $\bf 97.6$ & $\bf 97.7$ & $\bf 95.9$ \parbox{0pt}{\rule{0pt}{.5ex+\baselineskip}} \\ [0ex]  	
			
			\cline{1-19}
		\end{tabular}
	\end{center}
	\label{tab:performance}
	\vspace{-0.2cm}
\end{table*}

\section{Experiments}
\subsection{LITIS-Rouen dataset and modification for this study}
\label{ssec:dataset}
We conducted experiments using the LITIS-Rouen dataset \cite{Rakotomamonjy2015}. The dataset consists of 19 scene categories with 3026 examples in total. All instances have the same length of 30 seconds and were recorded with a sampling rate of 22050\,Hz. 

For this study, we did not use the full 30-second snippets. Instead, we decomposed each 30-second snippet into segments of 4 seconds length without overlap (except the last segment). Thus $S~=~8$ such segments were obtained from each 30-second snippet. 
The classification models were trained with 4-second segments extracted from the training data. 
To understand how the classification performance varies with different test signal lengths, we sequentially evaluated a classification model on the 4-second segments of a 30-second snippet and aggregated the classification results over $\{1, 2, \ldots, S\}$ segments which are equivalent to $\{4, 8, \ldots, 28, 30\}$ seconds. 
In order to aggregate classification results over multiple segments in the late-fusion system, we used the same fusion method that we used for model fusion. For the standalone CNN, RNN, and the two-stream C-RNN, we employed the probabilistical multiplication fusion of segment-wise results, as it has been shown to be efficient for this purpose \cite{Han2016,phan2017c}.

For an experiment with a specific test signal length, we follow the training/testing splits in the seminal work \cite{Rakotomamonjy2015} with the reported performances averaged over 20 splits.

\subsection{Network training and parameters}
\label{ssec:network_training}

A dropout rate of $0.5$ and $0.1$ was used for the CNN and the RNN, respectively. Particularly, for the early fusion C-RNN network, a dropout rate of $0.5$ was further applied to the fusion features. For all networks, the regularization parameter $\lambda$ was commonly set to $10^{-3}$ and the network training was accomplished using the \emph{Adam} optimizer \cite{Kingma2015} with a learning rate of $10^{-4}$. In addition, the hyper-parameter $C$ of the SVMs used for classification after network training was fixed to $1.0$.

\subsection{Experimental results}
\subsubsection{Overall performance}

\begin{figure*} [!t]
	\centering
	\includegraphics[width=0.95\linewidth]{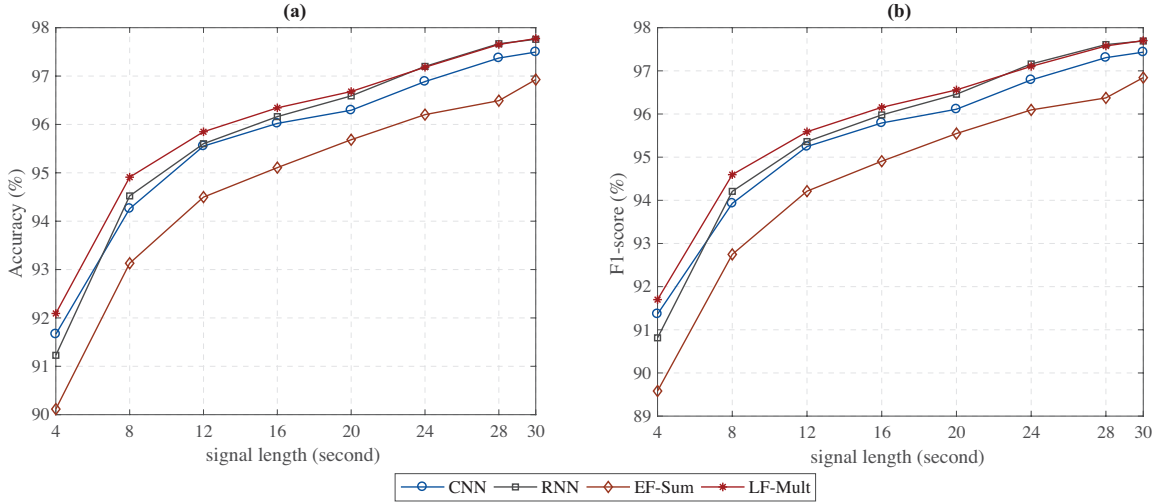}
	\caption{Variation in overall performance of CNN, RNN, EF-Sum (i.e. the best early fusion system) and LF-Mult (i.e. the best late fusion system) over different test signal lengths. (a) Accuracy and (b) F1-score.}
	\label{fig:overall_performance}
	\vspace{-0.1cm}
\end{figure*}

\begin{figure*} [!t]
	\centering
	\includegraphics[width=0.95\linewidth]{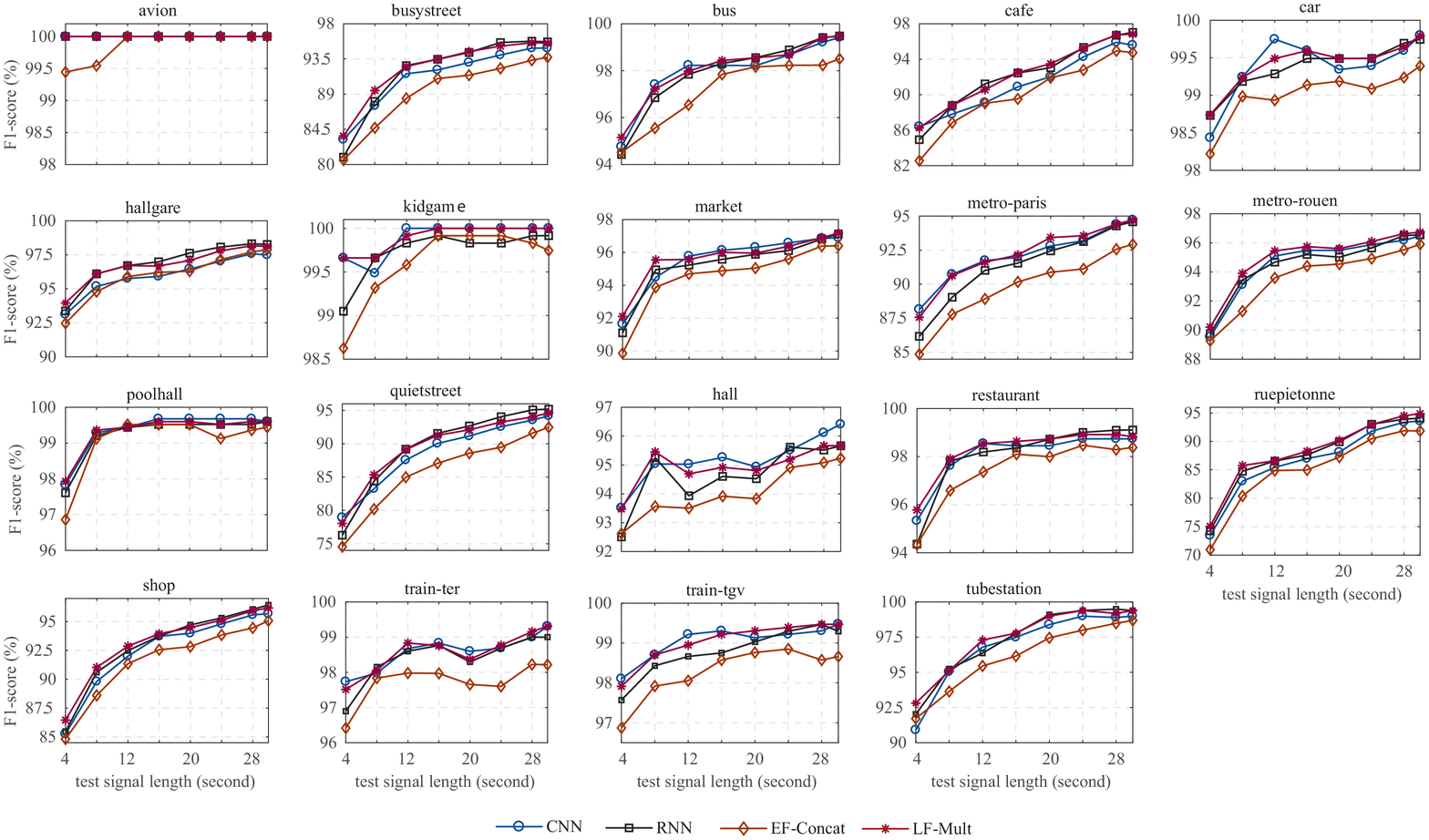}
	\caption{Variation in category-specific F1-scores of CNN, RNN, EF-Sum (i.e. the best early fusion system) and LF-Mult (i.e. the best late fusion system) over the range of test signal lengths.}
	\label{fig:specific_performance}
\end{figure*}

Table \ref{tab:performance} shows the overall performance in terms of accuracy and F1-score obtained by the studied classification systems over different test signal lengths $\{4,8,\dots,28,30\}$. Fig. \ref{fig:overall_performance} provides a visual snapshot of  
variation of four systems, including the CNN, RNN, the best early-fusion system (i.e. EF-Sum) and  and the best late-fusion system (i.e. LF-Mult), over the spectrum of the test signal length.

On average, the standalone CNN marginally underperforms its RNN counterpart. 
Their performance gap is particularly noticeable with large test signal lengths ($\ge 20$ seconds). These results highlight the importance of sequential modelling for temporal data. 
It may also be worthwhile noticing that the CNN proposed here achieves an F1-score of $97.4\%$, improving that of the over-time-convolution CNN in \cite{phan2017b} (i.e. $96.5\%$) by $0.9\%$ absolute. 
Regarding the fusion systems, sum fusion and concatenation fusion perform comparably in early fusion and appear to be better than max fusion. However, among the late fusion methods, multiplication fusion performs the best. 
Compared to other methods, multiplication fusion favours likelihood for categories that have consistent classification results, and more aggressively suppresses those with diverged classification results~\cite{Han2016, phan2017c}.

More importantly, the results in Table \ref{tab:performance} and Fig. \ref{fig:overall_performance} show that early fusion consistently worsens the classification performance. On average, the best early-fusion system (i.e. EF-Sum) reduces the F1-score by $1.0\%$ and $1.2\%$ absolute compared to the standalone CNN and RNN, respectively. 
Late fusion is more efficient as LF-Mult yields performance gains of $0.4\%$ and $0.2\%$ absolute on the average F1-score compared to the standalone CNN and RNN, which was previously reported with state-of-the-art performance at 30-second test signal length on the experimental dataset \cite{phan2017c}. 

\subsubsection{When is model fusion useful?}	

Inspection on Fig. \ref{fig:overall_performance} further reveals that model fusion is the most productive when the test signal length is small (i.e. $< 20$ seconds). 
When the test signal length is larger than 20 seconds, the performance gain (if any) achieved by LF-Mult is very marginal. Intuitively, with short signal lengths, multiple views on a short duration scene can compensate each other in the ensemble. 
However, when listening longer, the best standalone system (namely RNN) has accumulated more information about a scene and can recognize it quite reliably. 
Meanwhile, the weaker standalone model (namely the CNN) does not appear to bring sufficient new information about the scene into the ensemble.

\subsubsection{How long is sufficient to recognize a scene?}	

Fig. \ref{fig:specific_performance} shows the F1-score variation patterns for several different scene categories over the spectrum of test signal length. 
It is unsurprising to find that the patterns are very category-specific. Several categories can be reliably recognized even with a very short signal length. For example, we achieve perfect or near perfect recognition accuracy rate on ``avion'', ``kidgame'', and ``poolhall'' within 4, 12, and 16 seconds, respectively. 
Some other scenes, such as ``car'', ``metro-rouen'', ``restaurant'', ``train-ter'', and ``train-tgv'', can also be recognized reliably within 16 seconds, and accumulating more information about these scenes further improve performance.
In contrast, scenes such as ``cafe'', ``quitestreet'', ``ruepietonne'', and ``shop'' require much longer test signal lenghts (e.g. 30 seconds) to achieve good performance.
It would be expected that a system should listen even longer (i.e. $> 30$ seconds) to improve the detection certainty for those scenes. 

\section{Future research}

In this study, we made use of high-level LTE features that were learned to encode the hierarchy of the scene categories \cite{phan2016d} to train the networks. Future research should look for an alternative method to encode this hierarchy with a  neural network. Such a network should be trained jointly with the classification task in an end-to-end fashion. In addition, the networks were trained with audio segments of 4 seconds long and a coarse temporal resolution for signal decomposition and feature extraction (i.e. 250 ms frames) may overlook the fine structure of the signal. Using low-level features, such as log Mel-scale spectrograms \cite{Han2016}, and employing smaller frames for signal decomposition may avoid this potential issue. They would also allow us to train a network with shorter audio segments, enabling studying audio scene recognition with variable lengths at a finer temporal resolution. 

\section{Conclusion}
This work studied an important aspect of the audio scene recognition task, namely how much time is sufficient to recognize an audio scene. We additionally investigated which kind of model fusion and to which temporal extent model fusion is most useful for the task.
These studies were accomplished by exploring ASC using a CNN, an RNN and their fusion using various early and late fusion strategies.  
Experimental results on the LITIS-Rouen dataset showed that the duration required to recognize a scene is category-specific, just as it is for humans. 
Some scenes were reliably recognized within a few seconds while others required significantly longer durations.
Furthermore, while late fusion was shown to improve the classification performance, it is most beneficial when the test signal length is limited. These findings suggest that practical implementation of audio scene recognition systems should adapt their recognition strategies to different kinds of target scenes to improve their service quality.

\bibliographystyle{jaes}

\bibliography{reference}

\end{document}